\documentclass[conference]{IEEEtran}
\IEEEoverridecommandlockouts
\usepackage{listings}
\usepackage{xcolor}
\usepackage[nocompress]{cite}
\definecolor{codegreen}{rgb}{0,0.6,0}
\definecolor{codegray}{rgb}{0.5,0.5,0.5}
\definecolor{codepurple}{rgb}{0.58,0,0.82}
\definecolor{backcolour}{rgb}{0.95,0.95,0.95}
\lstdefinestyle{mystyle}{
    backgroundcolor=\color{backcolour},   
    commentstyle=\color{codegreen},
    keywordstyle=\color{magenta},
    numberstyle=\tiny\color{codegray},
    stringstyle=\color{codepurple},
    basicstyle=\ttfamily\footnotesize\bfseries,
    breakatwhitespace=false,         
    breaklines=true,                 
    captionpos=b,                    
    keepspaces=true,                 
    numbers=none,                    
    numbersep=5pt,                  
    showspaces=false,                
    showstringspaces=false,
    showtabs=false,                  
    tabsize=2
}
\usepackage{amsmath,amssymb,amsfonts}\usepackage{algorithmic}
\usepackage{graphicx}
\usepackage{textcomp}
\usepackage{xcolor}
\usepackage{authblk}
\usepackage{url}
\usepackage{natbib}
\def\BibTeX{{\rm B\kern-.05em{\sc i\kern-.025em b}\kern-.08em
    T\kern-.1667em\lower.7ex\hbox{E}\kern-.125emX}}
\begin{document}

\title{PatentsView-Evaluation: Evaluation Datasets and Tools to Advance Research on Inventor Name Disambiguation}

\author[1,2]{Olivier Binette} 
\author[1]{Sarvo Madhavan}
\author[1]{Jack Butler}
\author[1]{Beth Anne Card}
\author[1]{Emily Melluso}
\author[1]{Christina Jones}
\affil[1]{Duke University}
\affil[2]{American Institutes for Research}

\maketitle

\begin{figure*}
    \centering
    \includegraphics[width=0.8\linewidth]{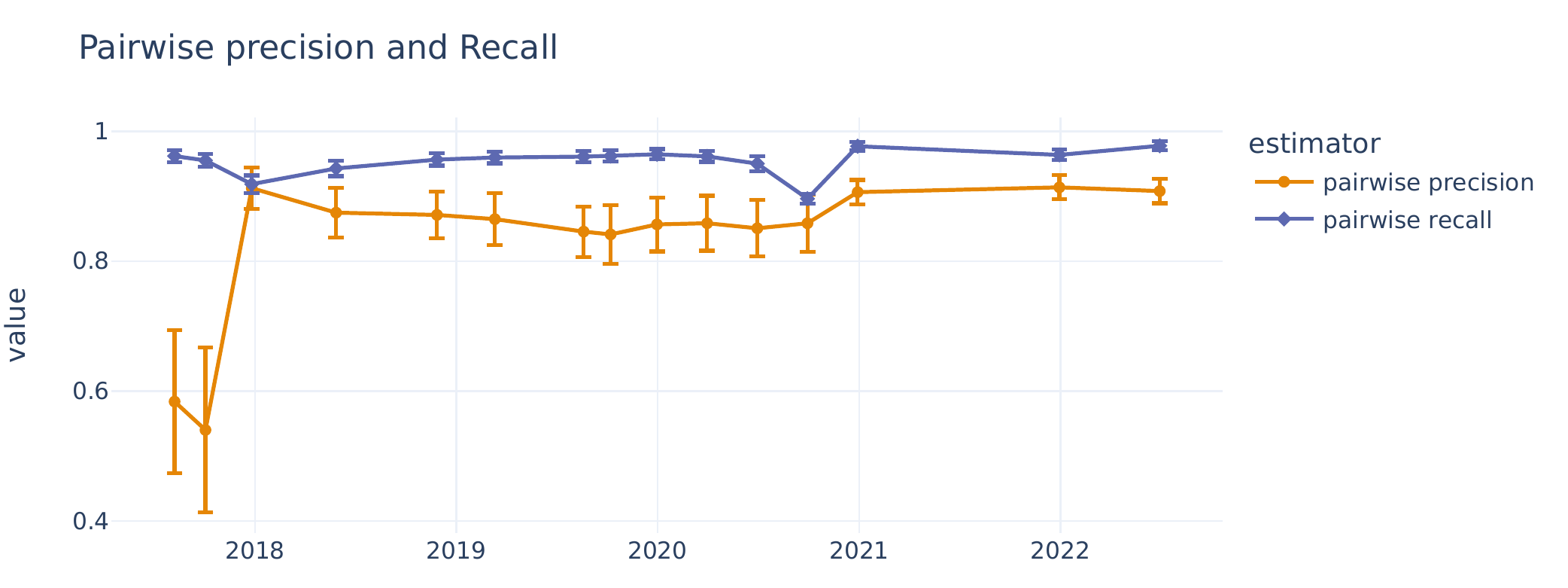}
    \caption{Pairwise precision and recall estimates over PatentsView's disambiguation history.}
    \label{fig:performance_estimates}
\end{figure*}


\begin{abstract}
We present PatentsView-Evaluation, a Python package that enables researchers to evaluate the performance of inventor name disambiguation systems such as PatentsView.org. The package includes benchmark datasets and evaluation tools, and aims to advance research on inventor name disambiguation by providing access to high-quality evaluation data and improving evaluation standards.
\end{abstract}

\begin{IEEEkeywords}
Digital libraries, Inventor name disambiguation, PatentsView, Statistical Evaluation, Open-source software
\end{IEEEkeywords}

\section{Introduction}
Inventor name disambiguation is the task of identifying unique inventors in patent datasets \citep{Li2014, toole2021patentsview}. This requires using contextual information to distinguish between different inventors with the same name and to resolve name variations. Since there are no unique identifiers for inventors on U.S. patents, disambiguation is done using statistical algorithms which provide approximate solutions. The task is closely related to author name disambiguation in digital libraries \citep{smalheiser2009author, ferreira2012brief, subramanian2021s2and} and is a particular case of entity resolution \citep{Christen2012, Christophides2019, Binette2022a}.

Unfortunately, progress in the field has been hindered by misleading evaluation methodology and a lack of representative benchmark datasets \citep{Wang2022}. Naively computing performance metrics (i.e., precision and F-score) on benchmark datasets leads to biased estimates and flipped rankings of competing algorithms in many cases \citep{Binette2022b}. {This is due to the non-trivial scaling of entity resolution performance: while it is easy to disambiguate small benchmark datasets, the opportunity for error grows quadratically as a function of dataset size.} Furthermore, some benchmark datasets are outdated or unavailable to the general public.

To address these challenges, we have released PatentsView-Evaluation, a Python package that is available at \url{github.com/patentsView/patentsView-Evaluation/} and that can be installed from PyPI \citep{pypi} using:
\begin{lstlisting}
pip install pv-evaluation
\end{lstlisting}
This is an open-source Python package which contains a suite of benchmark datasets and evaluation tools for representative performance evaluation. The package includes datasets used in the U.S. Patents and Trademarks Office (USPTO) 2015 disambiguation competition, Azoulay's Academic Life Sciences dataset which was previously unavailable to the general public, as well as a novel dataset {extending what was developed by PatentsView in \cite{Binette2022b}} specifically for evaluation purposes. To facilitate performance evaluation, the package also includes representative precision and recall estimators as well as a suite of summary statistics and visualizations.

The rest of the paper is structured as follows. In section \ref{sec:overview}, we provide an overview of the package's modules, including the available data and performance estimators. Section \ref{sec:discussion} summarizes our contributions and outlines our vision for future research.

\section{Overview of the Package}\label{sec:overview}

PatentsView-Evaluation is built on top of the ER-evaluation Python package \citep{Binette2022c} which provides its core entity resolution evaluation functionality. It contains two main submodules. The \texttt{benchmark} module provides data, summary statistics, and visualizations. The \texttt{template} module provides templated reports that can be compiled to html using the Quarto publishing system (quarto.org).

\subsection{Benchmark Datasets}

Inventor disambiguation associates \textbf{inventor mentions} to unique inventor identifiers. Here, an inventor mention is the combination of a patent number and an authorship sequence number, resulting in a \textbf{mention ID}. For instance, the mention ID ``US11379060-0" refers to the first inventor listed on U.S. patent number 11379060. 

Our benchmark datasets are pandas Series \citep{reback2020pandas} indexed by inventor mentions and with values corresponding to a unique inventor identifier. Note that, while benchmark datasets aim to provide a ground truth disambiguation of a set of inventors, they may still contain errors resulting from the inherent uncertainty and difficulty of disambiguating inventors.

The inventors benchmarks which we provide are listed below. These are available in the package through functions named \texttt{load\_*\_inventors\_benchmark()}. 



\begin{enumerate}
    \item The \textbf{Academic Life Sciences (ALS)} dataset from the file named ``patents\_2005\_12" was graciously shared by Pierre Azoulay (personal communication) with permission to release the corresponding clustering of inventor mentions. This dataset and variations of it were referred to in \cite{azoulay2007determinants, Azoulay2011, Ventura2015}. We prepared the data by associating mention IDs to each record based on patent numbers and inventor mention names.
    \item The \textbf{Israeli inventors benchmark} from \cite{trajtenberg2008identification}.
    \item \textbf{Li's 2011 inventors benchmark} from \cite{Li2014}.
    \item The \textbf{Engineer and Scientist inventors benchmark} from PatentsView's 2015 disambiguation competition \citep{PatentsView2015}.
    \item \textbf{PatentsView's 2021 inventors benchmark} from \cite{Monath2021}, which contains a set of particularly ambiguous inventor mentions.
    \item \textbf{Binette's 2022 inventors benchmark} which extends \cite{Binette2022b} and covers U.S. patents granted between 1976 and December 31, 2021. This is a random sample of inventors with sampling probabilities proportional to an inventor's number of patents.
\end{enumerate}

\subsection{Performance Estimators}

As previously noted, naively computing precision and recall on benchmark datasets results in misleading figures. As such, PatentsView-Evaluation borrows from \cite{Binette2022b} methodology for representative performance estimation. Given a set of inventor disambiguations for U.S. patents granted between 1976 and December 31, 2021, the function \texttt{inventor\_estimates\_trend\_plot()} provides a plot of estimated precision and recall for each disambiguation with uncertainty quantification ($\pm$ one standard deviation). By default, these estimates are based on Binette's 2022 inventors benchmark. Estimates corresponding to the use of other benchmark datasets can be obtained by passing them as additional arguments. Figure \ref{fig:performance_estimates} showcases the resulting plot with default arguments.

\begin{figure*}
    \centering
    \includegraphics[width=0.8\linewidth]{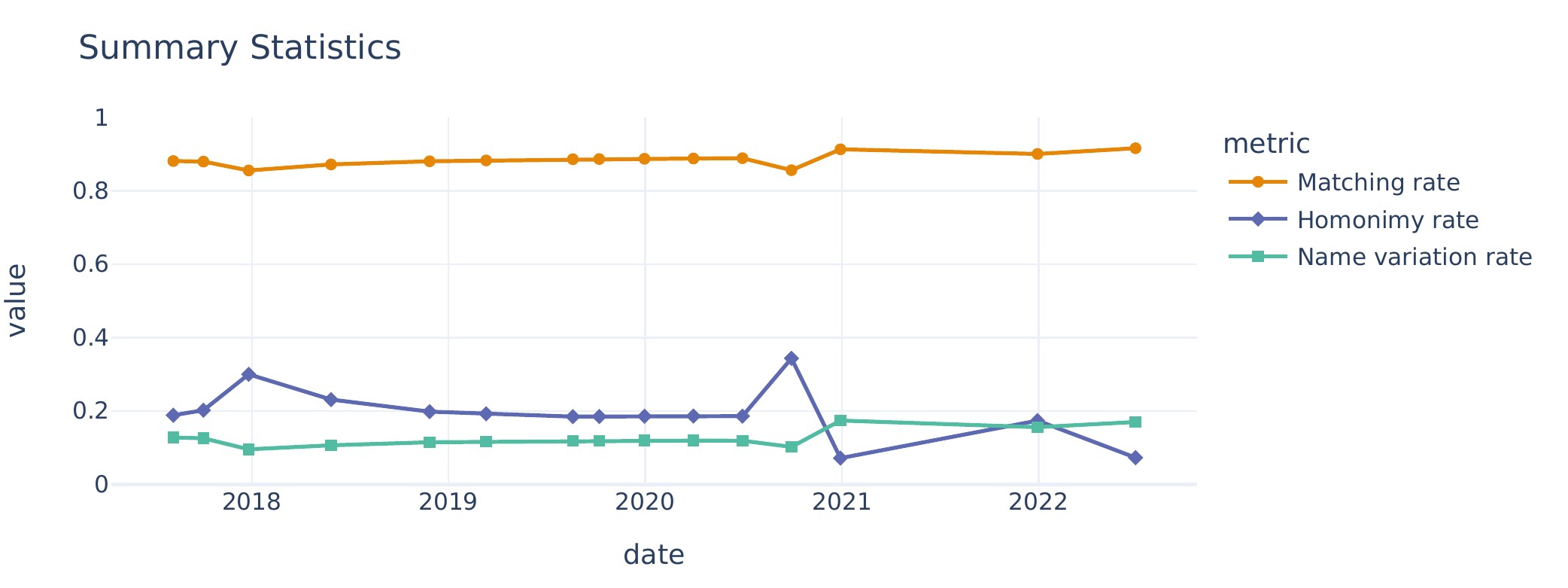}
    \caption{Evolution of summary statistics over PatentsView's disambiguation history.}
    \label{fig:metrics}
\end{figure*}

\subsection{Summary Statistics and Visualizations}

In addition to performance metric estimators, PatentsView-Evaluation provides a suite of summary statistics visualizations based on the ER-Evaluation package. This allows monitoring metrics such as the matching rate, the name variation rate, name homonymy rate, and the cluster size distribution entropy. More information on the definition of these metrics is provided in \cite{Binette2022c}. The function \texttt{inventor\_summary\_trend\_plot()} provides one entry point to visualizing these metrics for PatentsView's disambiguation history. Figure \ref{fig:metrics} showcases its output. {Notice how, around 2021, the homonymy rate changes from around $0.2$ to nearly $0.4$ before going back down close to $0.05$. These are major differences to the disambiguation which are not reflected in the matching rate.}

\subsection{Templated HTML Reports}

The last component of PatentsView-Evaluation is a templated report which can be compiled to HTML using Quarto. It allows the comparison of a set of inventor disambiguations and through summary statistics, evaluation metrics, and error visualization. The entry point is the function \texttt{render\_inventor\_disambiguation\_report()} which takes as arguments a set of disambiguation files.

\section{Discussion}\label{sec:discussion}

In this paper, we presented PatentsView-Evaluation, a Python package with evaluation data and tools to advance inventor name disambiguation. We provided an overview of the package as well as a few examples of its capabilities.

PatentsView's vision for improved inventor name disambiguation builds upon its experience and the success of its existing system. We aim to improve the maintainability, modularity, and performance of PatentsView's system through separate innovation within its three main components: (1) the feature engineering component which defines pairwise comparison metrics for given patent attributes, (2) the similarity modeling component which estimates pairwise match probabilities, and (3) the clustering component which resolves transitive inventor clusters. For (1), we aim to develop additional features through the use of modern text analysis and natural language processing methods. For (2), we aim to develop flexible semi-supervised methods which can account for dependencies between features and biases in the benchmark datasets. Finally, for (3), we aim to better tune clustering algorithms to optimize key performance metrics. Through the use of principled performance evaluation tools available in the PatentsView-Evaluation package, new methodological developments can now be rigorously tested.


\bibliographystyle{chicago}
\bibliography{main}

\end{document}